\documentclass[conference]{IEEEtran}
\IEEEoverridecommandlockouts
\usepackage{cite}
\usepackage{amsmath,amssymb,amsfonts}
\usepackage{algorithm}
\usepackage{algorithmicx}
\usepackage{algpseudocode}
\usepackage{graphicx}
\usepackage{textcomp}
\usepackage{xcolor}
\usepackage{amsmath,epsfig}
\usepackage{multirow}
\usepackage{array}
\usepackage{graphicx}
\usepackage{subfigure}
\usepackage{booktabs}
\usepackage{bm}
\def\BibTeX{{\rm B\kern-.05em{\sc i\kern-.025em b}\kern-.08em
    T\kern-.1667em\lower.7ex\hbox{E}\kern-.125emX}}
\begin{document}

\title{Adaptive Few-Shot Learning Algorithm for Rare Sound Event Detection
	\thanks{${* }$ Corresponding Author: Jianzong Wang, jzwang@188.com}}

\author{\IEEEauthorblockN{Chendong Zhao$^{\dagger \star \ddagger}$\thanks{$\ddagger$ Work done during an internship at Ping An Technology. }, Jianzong Wang$^{\dagger * }$, Leilai Li$^{\dagger}$, Xiaoyang Qu$^{\dagger}$, Jing Xiao$^{\dagger}$}
	\IEEEauthorblockA{$^{\dagger}$ \textit{Ping An Technology (Shenzhen) Co., Ltd., Shenzhen, China} \\
		{$^{\star}$ \textit{The Shenzhen International Graduate School, Tsinghua University, China}}
}}

\maketitle

\begin{abstract}
Sound event detection is to infer the event by understanding the surrounding environmental sounds.
Due to the scarcity of rare sound events, it becomes challenging for the well-trained detectors which have learned too much prior knowledge.   
Meanwhile, few-shot learning methods promise a good generalization ability when facing a new limited-data task. Recent approaches have achieved promising results in this field.
However, these approaches treat each support example independently, ignoring the information of other examples from the whole task. 
Because of this, most of previous methods are constrained to generate a same feature embedding for all test-time tasks, which is not adaptive to each inputted data. 
In this work, we propose a novel task-adaptive module which is easy to plant into any metric-based few-shot learning frameworks. The module could identify the task-relevant feature dimension. Incorporating our module improves the performance considerably on two datasets over baseline methods, especially for the transductive propagation network. Such as +6.8\% for \textit{5-way 1-shot} accuracy on ESC-50, and +5.9\% on noiseESC-50. We investigate our approach in the domain-mismatch setting and also achieve better results than previous methods.
\end{abstract}


\section{Introduction}
\label{sec:intro}

Automatic environmental sound event detection has received increasing attention in recent years~\cite{TUTDatatbase,r}. It deals with audios detecting and classifying, which leads to multi-form applications in industry. Environmental sound is naturally different from other audios. It doesn't exhibit any stationary patterns like phoneme in speech~\cite{fed,g2p} or rhythm in music. In contrast, sound event contains very complex temporal structure~\cite{sed1,sed2,se} that may be continues (e.g. rains), abrupt (e.g. thunder storm) or periodic (e.g. clock tick). Moreover, speech~\cite{5,6} and music usually distribute on a relatively fixed frequency bandwidth, but sound event spans a wide frequency range where different sound’s frequency may distribute in different range that leads to various patterns, such as the ”Airplane” and ”Dog” sounds' frequency differs a lot. The information contained in temporal patterns and frequency bins of the sound event could be massive~\cite{3}.

\begin{figure}[t]
	\centering
	\includegraphics[width=\linewidth]{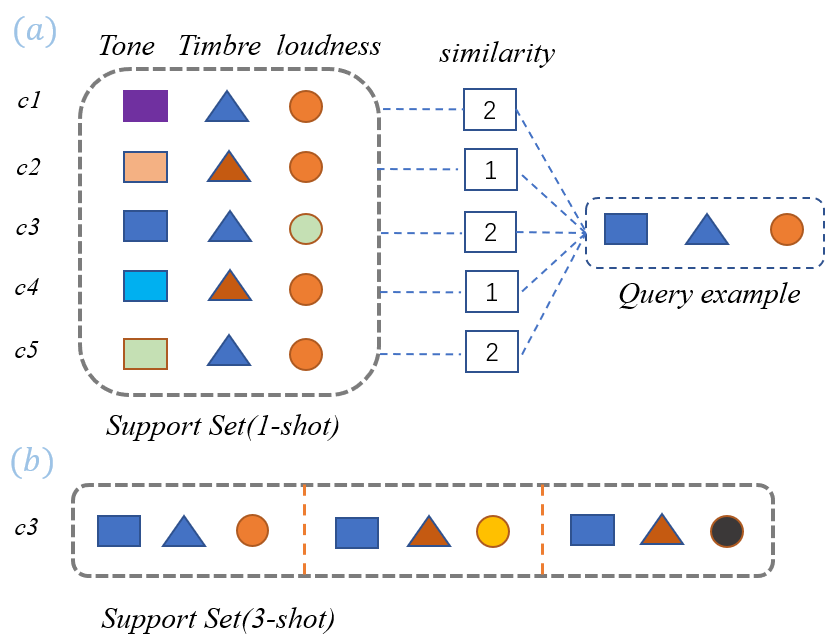}
	\caption{An toy example to illustrate the motivation. (a) defines a 5-way 1-shot task. There’re three features: \textit{Tone}, \textit{Timbre} and \textit{loudness}. Different color means different value of the feature, same color adds the similarity score by 1; (b) In the k-shot (k=3) setting, all examples of class c3 share the same value of \textit{Tone} even though their \textit{Timbre} and \textit{loudness} are different. }
	\label{fig:idea}
\end{figure}

To date, many deep-learning (DL) methods greatly improved detection performance~\cite{autoTagging,2,audiofew,song,park,few-shot-sound-detection}. However, they typically require large amounts of labeled data, which limits the generalization ability to limited-data tasks due to the annotation cost. In contrast, children can recognize new things quickly with proper guidance, even they only see the examples few times. These motivate the study of \textbf{Few-shot learning}. Meantime, this method has also been introduced in~\cite{attentionSimilarity,rare,few-att-gnn,fewShotGNN,TPN} which concerning rare sound event detection and achieved promising results. Few-shot model promises alleviating the problem of data deficit and cold start. It usually follows the episodic training strategy~\cite{attentionSimilarity,MatchNet}, which considers an $N$-way $K$-shot (e.g 5 way, 1 shot means there’re 5 classes, each contains 1 support example) classification task in each episode.

In the few-shot learning setting, a model is first trained on labeled data with base classes. Then, model generalization is evaluated on few-shot tasks, composed of unlabeled samples from novel classes unseen during training (query set), assuming only one or a few labeled samples (support set) are given per novel class.
All metric-based few-shot learning frameworks~\cite{fewShotGNN, TPN, protoNet, relationNet, CloserLook} compute similarity between each support (training) example and the query example independently, resulting in the correlation among support examples being missed. Figure~\ref{fig:idea} illustrates our motivation of the task-adaptive module. The goal of
this task is to correctly classify the query example. In Figure~\ref{fig:idea} (a), we could notice that each support example has different \textit{Tone}, but may have same \textit{Timbre} and \textit{loudness} with others. During the similarity computation, the score between support example (c1, c3, c5) and the query example are all 2, making it hard to classify the query example. Moreover, in multi-shot setting like Figure~\ref{fig:idea} (b) shows, most of class c3 have the same \textit{Tone} but various \textit{Timbre} and \textit{loudness}. So it's obvious that the critical feature in this task should be \textit{Tone} thus the label of the query example should be c3. Such an phenomenon occurs especially when $K$ is low, which motivates us incorporating the context of all categories in each task to find the relevant features. Our task-adaptive module aims to integrate all support examples' information to value the commonality within per class and the uniqueness among all classes, thus to find the critical features. 

Our contributions are as follows:

(1) We introduce a feature encoder integrating attention mechanism to capture the temporal\&channel context. In addition, a sound event-oriented data augmentaion strategy is introduced to alleviate the data-shortage problem. 

(2) We extend metric-based few-shot learning frameworks with a task-adaptive module to identify the uniqueness among classes and the commonality within per class of the whole support set. The output of
this module combines both support set and query set, making the subsequent feature embedding more effective.

(3) We evaluated our model on two benchmarks. Results verify the superiority of our model over previous methods. Besides, our model also achieves better performance in the setting of domain mismatch.

\section{Preliminary and related works}
\subsection{Few-shot sound event detection}
Recent approaches~\cite{audiofew,attentionSimilarity,few-att-gnn} adopt the prototypical networks~\cite{protoNet} and graph neural networks~\cite{fewgnn} for few-shot sound event detection. Few-shot sound event detection aims to correctly classify unseen sounds with a few support labeled examples to finetune the model.

Few-shot learning follows the episodic training paradigm that used in previous literature \cite{protoNet,maml}. Supposing there're two non-overlapping
datasets of classes (events) $\bm{C}_{train}$ and $\bm{C}_{test}$, where $\bm{C}_{train} \cap \bm{C}_{test} = \emptyset$. There are also two procedures: Meta-training and Meta-testing. In each episode of training, we randomly sample $N$ classes (a small subset) from $\bm{C}_{train}$ to construct support set $\mathcal{S}$ and query set $\mathcal{Q}$. A simple $N$-way $K$-shot task denotes as follow: $\mathcal{S}$ is denoted as $\mathcal{S} = \{x_1^1,...,x_K^1,..,x^N_1,..,x^N_K\}$, where $K$ is the number of samples in per class. The query set $\mathcal{Q} = \{ x_1^*,...,x_T^*\}$ contains various examples from the same $N$ classes. Thus, there are $N * K$ samples in $\mathcal{S}$ and $T$ samples in $\mathcal{Q}$.  The support set and query set composed a multi-label classification task here. During the procedure of meta-testing, $\mathcal{S}$ and $\mathcal{Q}$ sampled from $\bm{C}_{test}$ , and few-shot method is required to predict on query set (no label) given the support set (with label).

\begin{figure}[t]
	\centering
	\includegraphics[width=\linewidth]{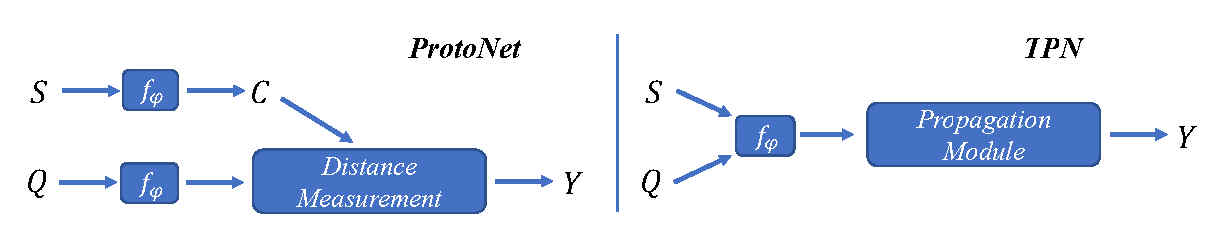}
	\caption{ Two metric-based few-shot learning methods. }
	\label{fig:fewshot}
\end{figure}

\subsection{Metric-based learning methods}
\label{methods}
Few-shot learning methods can be divided into three branches: metric-based, optimization-based and data augmentation-based~\cite{opt1,opt2,koch2015siamese}. In this paper, we mainly focus on metric-based learning. The metric-based methods could be categorized into inductive and transductive, and two representative methods are prototypical network~\cite{protoNet} and transductive propagation network~\cite{TPN}, as shown in Figure~\ref{fig:fewshot}.
Given a $N$-way $K$-shot task $\mathcal{T} = (\mathcal{S}, \mathcal{Q})$, and the feature encoder $f_\varphi$ decided by its internal parameter $\varphi$.

\noindent\textbf{Prototypical network} (\textit{ProtoNet}): this approach simply integrate nerual network (NN) baseline into the end-to-end meta-learning framework. It takes the average of the learned representations of a few examples for each class as class-wise representations $C$, and then classifies an unlabeled instance
by calculating the Euclidean distance between the input and the class-wise representations. $x^s_i \in \mathcal{S}$ and $x^q \in \mathcal{Q}$, and $\mathcal{M}$ is a pair-wise feature distance function. The distance measure is as follows:
\begin{equation}
	f_{sim}(x^s, x^q;\mathcal{S},\mathcal{Q},\varphi) =  \mathcal{M}(  \frac{1}{K}\sum_{i=1}^{K}f_\varphi(x^s_i),f_\varphi(x^q))
\end{equation}

\noindent\textbf{Transductive propagation network} (\textit{TPN}): this approach utilizes the entire test set for transductive inference, which is to consider the relationships among testset and thus predict them as a whole. Transductive inference has shown to outperform inductive methods~\cite{TSVM,TPNzero,cvpr}. TPN propose to learn to propagate labels via episodic paradigm. During the propagation, a distance measurement and example-wise length scale parameter were adopted to obtain a proper neighborhood graph. After the graph construction, label propagation determines the labels of the query set. The distance measure is as follows:
\begin{equation}
	f_{sim}(x_i, x_j;\mathcal{S},\mathcal{Q},\varphi,\phi) = exp( -\frac{1}{2}\mathcal{M}(\frac{f(x_i)}{\sigma_{i}} ,\frac{f(x_j)}{\sigma_{j}}))
\end{equation}
where $\phi$ is the parameters generating example-wise length-scale parameter ($\sigma_{i}, \sigma_{j}$). $x_i,x_j \in \mathcal{S} \cup \mathcal{Q}$.

\begin{figure*}[t]
	\centering
	\includegraphics[width=\linewidth]{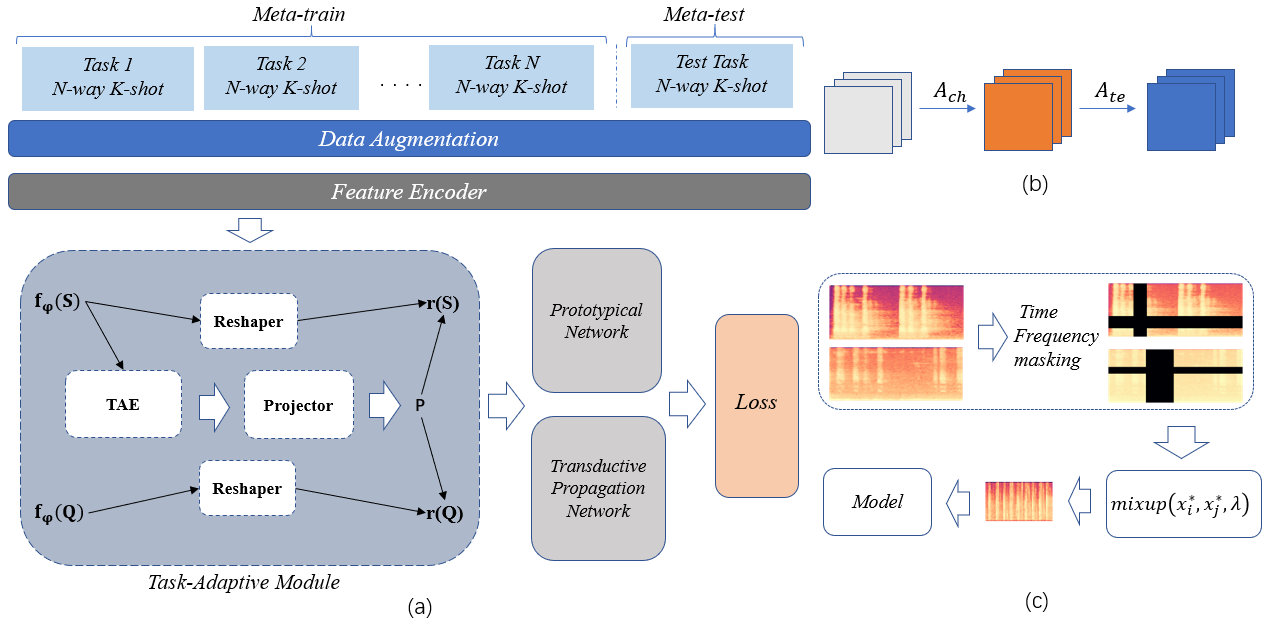}
	\caption{(a). The overall framework of our model, it is composed of three parts: feature encoder, task-adaptive module and metric-based few-shot learning network. (b). The temporal\&channel attention mechanism. $A_{ch}$ is the channel attention, $A_{te}$ is the temporal attention. (c). Data augmentation pipeline for the input log mel-spectrogram.}
	\label{fig:model}
\end{figure*}
\section{Approach}
The overview of the proposed learning framework is shown in Figure 3. In this section, we first introduce how to mask and mixup for augmenting the training data, then use class prototypes to build a classifier and update the classifier with transductive inference. After that, we discuss how to make use of the intergreted class prototypes to fine-tune the feature extractor. Lastly, we summarize the core idea of task-aware learning framework.
\subsection{Masked mixup}
To avoid possible overfitting caused by limited trained data, we adopt time and frequency masking~\cite{timefrequnencymask} and mix-up as the data augmentation strategy, which is simple but effective~\cite{dataa}. The strategy adopt multiple time and frequency masks on input spectrogram to generate multi-masked spectrograms and then randomly mixes two masked samples, increasing the diversity of samples by the way.
As shown in Figure~\ref{fig:model} (c), given a spectrogram $x$ with $T$ frames and $F$ frequency bins. The first step is to use multiple time masking and frequency masking, generating $M$ masked samples $x^{*}$.
To be specific, $Mask_{num}$ assigns $t$ consecutive time frames $[t_0:t_0+t)$ and $f$ consecutive frequency bins $[f_0:f_0+f)$ value to 0, and $num$ denotes the number of multiple masking. $t$ is chosen from a uniform distribution from 0 to the time parameter $\tau$, $t_0$ is chosen from $[0,T-t)$, $f$ is chosen from a uniform distribution from 0 to the frequency parameter $\upsilon$, $f_0$ is chosen from $[0, F-f)$.
Secondly, the mixup step conducts a convex combination of two randomly selected $(x_i^{*}, y_i)$ and $(x_j^{*}, y_j)$ from all masked samples:
\begin{align}
	&x^{*} \leftarrow x \odot Mask_{num} \\
	&\tilde{x} = \lambda x_i^{*} + (1 - \lambda)x_j^{*} \\
	&\tilde{y} = \lambda y_i + (1 - \lambda)y_j 
\end{align}
where $y_i$ and $y_j$ are one-hot encoded class labels, $(\tilde{x}, \tilde{y})$ being the new sample. $\lambda \in [0,1]$ is acquired by sampling from a beta distribution $Beta$($\alpha, \alpha$) with $\alpha$ being a hyperparameter.
\noindent\textbf{Feature encoder:} After data augmentation, the samples first encoded by a ConvNet $f_{\varphi}$ as same as~\cite{attentionSimilarity}, but the CNN layer replaced with a temporal\&channel attention CNN layer~\cite{timefrequnencymask} as shown in Figure~\ref{fig:model} (b). The architecture of ConvNet contains five blocks and a fully-connected layer, and the first two block includes a $3 \times 3$ convolutional layer, a batch normalization layer, a ReLU activation layer and a $4 \times 4$ max-pooling layer, and the last block is same as the former except the last $1\times1$ max-pooling layer.

\subsection{Task-adaptive module}
This module contains three parts and leverages the $f_{\varphi}(.)$ as input, and outputs the task-adapted feature embedding, which will be passed to subsequent metric-based learning network.

\noindent\textbf{Task-adaptive-extractor for commonality among classes:}
Task-adaptive-extractor (TAE) aims to find the commonality among all instances within a class. Denote the output shape from feature encoder $f_{\varphi}$ as $(N\times K, m_1, w_1, h_1)$, where $m_1, w_1, h_1$ indicate the number of channel and spatial size respectively. \textit{TAE} is defined as follows:
\begin{equation}
	f_{\varphi}(\mathcal{S}) : (N \times K, m_1, w_1, h_1) \stackrel{TAE}{\longrightarrow} o: (N, m_2, w_2, h_2)
\end{equation}
where $m_2, w_2, h_2$ denote the output number of channel and spatial size. In this part, we first utilize a simple CNN layer to perform the dimension reduction. Then the samples in each class are averaged to a final output $o$. The purpose of \textit{TAE} is to extract the commonality among a category. Specifically, for 1-shot setting, there is no average operation.
The purpose of \textit{TAE} is to eliminate the differences among instances and extract the commonalities in the same category.

We intend to make use of $o$ denoting the mean of the learned representation of the support samples on the support set to represent the learned representation of query samples. $N$ denotes the number of query samples. In our experiments, $sim$ stands for cosine similarity. We do not use $M$ for the reason that query sample is randomly chosen. 

\noindent\textbf{Projector for characteristics among classes:}
The goal of the second component namely projector is to find the characteristics of various classes. Projector takes the output of \textit{TAE} as input and produce a mask for the support and query set by observing all the support classes at the same time.
\begin{equation}
	o: (N, m_2, w_2, h_2) \stackrel{reshape}{\longrightarrow} \hat{o} \stackrel{CNN}{\longrightarrow} p:(1, m_3, w_3, h_3)
\end{equation}
During the projector process, firstly, we reshape the $o:(N, m_2, w_2, h_2)$ into $\hat{o}: (1, N \times m_2,w_2,h_2)$, then a small CNN is applied to $\hat{o}$, producing the mask $p: (1, m_3, w_3, h_3)$. Finally, a \textit{softmax} is also applied to the dimension $m3$.
For making the output of projector $p$ influence the feature encoder output $f_{\varphi}(.)$, we need to match the shape between $p$ and $f_{\varphi}(.)$. This could be achieved as follows: 
\begin{equation}
	f_{\varphi}(.) \stackrel{Reshaper}{\longrightarrow} r(.): (N\times K, m_3, w_3, h_3)
\end{equation}
where \textit{Reshaper} means a light-weight CNN network, and $r(.)$ is regarded as the \textit{Reshaper} network.

This idea is inspired by~\cite{m}, which aims to maximize the mutual information between the query features and their label predictions for a few-shot task at inference. It means that the model has seen these unlabeled data before making final prediction.

\noindent\textbf{Portable to metric-based backbone:}
The task-adaptive module is portable, which could be easily integrated with any metric-based few-shot learning methods. In this paper, we investigate two classical metric-based methods: prototypical network (inductive) and transductive propagation network (transductive) in Sec~\ref{methods}, both do not consider the whole support set at the same time. 
For support set, mask $p$ directly onto the embedding. For the query set, $\odot$ stands for broadcasting the value of $p$ along the sample dimension ($NK$) in $\mathcal{Q}$. So the distance measurement could be modified as follows. Specifically, $\theta$ is a model parameter in TPN.
\begin{equation}
	f_{sim}(\mathcal{S},\mathcal{Q},\varphi, \theta) = \mathcal{M}(p \odot  r(f_\varphi(\mathcal{S})), p \odot r(f_\varphi(\mathcal{Q})))
\end{equation}
Our loss function is based on the cross entropy following most of state-of-the-art methods:
\begin{equation}
	\mathcal{L} = \frac{exp(\sum_j^{K}f_{sim}(x_j,x_q))}{\sum_{i=1}^{N}exp(\sum_j^Kf_{sim}( x_j^i,x_q))}
\end{equation}
where $x_j^i$ , $x_q$ denoting support and query examples respectively.

\section{Experiments}
\subsection{Experimental setting}
\noindent\textbf{Data preparation}: We utlize ESC-50 and noiseESC-50 datasets in this work, which is from DCASE2021 task 5~\cite{dd}. The ESC-50 dataset contains 2,000 5-seconds audio clips that belonged to 50 classes, each having 40 examples. The sound categories cover sounds of human behaviors (sneezing, crying, laughing), animals (bird, dog, sheep), machine sounds (train, clock-alarm, airplane). 
Our work also follows~\cite{attentionSimilarity} to evaluate the performance under noise condition called noiseESC-50 that selected from audio recordings of 15 different acoustic scenes. So, the performance on ESC-50 and noiseESC-50 would reflect the generalization ability of the model in real-world applications. 
What's more, although ESC-50 is relatively smaller than AudioSet, which suffers from the class imbalance problem, this also is the reason for our choice.
To directly compare our model with other baselines, we follow the setting of~\cite{MatchNet} as same as~\cite{attentionSimilarity}. Two datasets are divided into 35 classes for training, 10 classes for test and other classes for validation. All audio clips are down-sampled from 44.1kHz to 16kHz. 128-bin log mel-spectrogram of raw audio is extracted as the input.

\begin{table*}[t]
	\centering
	\setlength{\tabcolsep}{5pt}
	\caption{The result of sound detection (in \%) on ESC-50 and noiseESC-50. All baselines reported here are directly reprint the experimental results from the literature~\cite{attentionSimilarity}. \textbf{TA} means the task-adaptive module. \textbf{DA} means the data augmentation method.}
	\label{tab:1}
	\resizebox{1.0\linewidth}{34mm}{
		\begin{tabular}{c|c|c|c|c|c|c|c|c}
			\toprule[1.5pt]
			\multirow{3}{*}{~~~~~~Model~~~~~~} & \multicolumn{4}{c|}{~~~~~~ESC-50~~~~~~} & \multicolumn{4}{c}{~~~~~~noiseESC-50~~~~~~} \\ \cline{2-9} 
			& \multicolumn{2}{c|}{5-way acc} & \multicolumn{2}{c|}{10-way acc} & \multicolumn{2}{c|}{5-way acc} & \multicolumn{2}{c}{10-way acc} \\ \cline{2-9} 
			& {1-shot}     & {5-shot}   & {1-shot}   & {5-shot}   & {1-shot}   & {5-shot}  & {1-shot}   & {5-shot}  \\ \midrule
			{~~~MatchingNet \cite{MatchNet}~~~} & ~~~~53.7\%~~~~ & ~~~~67.0\%~~~~ & ~~~~34.5\%~~~~ & ~~~~47.9\%~~~~ & ~~~~51.0\%~~~~ & ~~~~61.5\%~~~~ & ~~~~31.7\%~~~~ & ~~~~43.0\%~~~~  \\ \midrule
			{RelationNet \cite{relationNet}} & 60.0\% & 70.3\% & 41.7\% & 52.0\% & 56.2\% & 74.5\% & 39.2\% & 52.5\% \\ \midrule
			{SimilarityEmbeddingNet \cite{SimilairtyEmbeddingNetwork}} & 61.0\% & 78.1\% & 45.2\% & 65.7\% & 63.2\% & 78.5\% & 44.2\% & 62.0\% \\ \midrule
			{ProtoNet \cite{protoNet}} & 67.9\% & 83.0\% & 46.2\% & 74.2\% & 66.2\% & 83.0\% & 46.5\% & 72.2\% \\ \midrule
			{ProtoNet + AS \cite{attentionSimilarity}} & 74.0\% & 87.7\% & 55.0\% & 76.5\% & 69.7\% & 85.7\% & 51.5\% & 73.5\% \\ \midrule
			{TPN \cite{TPN}} & 74.2\% & 86.9\% & 55.2\% & 76.7\% & 72.7\% & 86.1\% & 52.7\% & 72.9\% \\ \midrule \midrule
			{\textbf{\textit{ProtoNet+DA}}} & 70.5\%  & 83.3\%  &  48.9\%  & 74.7\% & 69.8\%  & 83.3\%  & 47.7\%  & 72.1\%  \\ \midrule
			{\textbf{\textit{TA+ProtoNet}}} & 70.2\%  & 84.0\%  &  48.6\%  & 74.8\% & 69.5\%  & 83.5\%  & 50.1\%  & 72.4\%  \\
			\midrule
			{\textbf{\textit{TA+ProtoNet+DA}}} & 70.8\%  & 84.6\%  & 50.9\%  & 75.3\% & 71.5\%  & 84.8\%  & 50.2\%  & 72.3\%  \\
			\midrule
			{\textbf{\textit{TA+ProtoNet+DA (temporal\&channel)}}} & 71.6\%  & 85.2\%  & 51.5\%  & 75.7\% & 72.1\%  & 85.2\%  & 51.3\%  & 72.9\%  \\
			\midrule
			{\textbf{\textit{TPN+DA}}} & 77.3\%  & 86.5\%  & 60.1\%  & 75.7\% & 77.1\% & 86.6\% & 55.9\%  & 73.1\%  \\
			\midrule
			{\textbf{\textit{TA+TPN}}} & 76.5\%  & 87.1\%  & 57.8\%  & \textbf{76.9}\% & 76.3\% & 86.3\% & 57.3\%  & 73.2\%  \\
			\midrule
			{\textbf{\textit{TA+TPN+DA}}} & 80.2\%  & 86.4\%  & 62.3\%  & 76.1\% & 77.9\%  & 86.8\%  & 59.3\%  & 76.1\%  \\
			\midrule
			{\textbf{\textit{TA+TPN+DA (temporal\&channel)}}} & \textbf{81.0\%}  & \textbf{87.2}\%  & \textbf{63.6}\%  & 76.8\% & \textbf{78.6\%}  & \textbf{87.1}\%  & \textbf{60.5}\%  & \textbf{76.7}\%  \\
			\bottomrule[1.5pt]
		\end{tabular}
	}
\end{table*}
\begin{table}[t]
	\caption{Architecture of ConvBlock layer of TAE and Encoder Architecture.}
	\label{tab:3}
	\centering
	\resizebox{1.0\linewidth}{11mm}{
		\begin{tabular}{cc|cc}
			\toprule
			\multicolumn{2}{c|}{ConvBlock Architecture} & \multicolumn{2}{c}{Encoder Architecture} \\ \hline
			Sub-Layer 1 & Conv2D &Layer 1 & ConvBlock 128         \\
			Sub-Layer 2 &  BatchNorm &Layer 2 & ConvBlock 128\\
			Sub-Layer 3 & ReLU &Layer 3 & ConvBlock 128 \\
			Sub-Layer 4 & MaxPool2D((2,2)) &Layer 4 & Flatten\\ \bottomrule
		\end{tabular}
	}
\end{table}

\noindent\textbf{Implentation detail}:  We list the model arcgitecture in Table~\ref{tab:3}. The librosa~\cite{librosa} is used for feature extraction. During the episodic training, for each task, we only perform the mixup on the query set $\mathcal{Q}$.  
Empirically, $\tau=24$, $\upsilon=36$ and $num=2$ are used for time and frequency masking, and $\alpha=0.2$  is used for masked mixup.
We conduct Adam optimization with stochastic gradient descent algorithm in all experiments with an initial learning rate of 0.0001. The model is trained by using back-propagation with cross entropy loss function as we mentioned before. The network is trained for a max of 60 epochs and a decaying factor 0.01 is set for the learning rate to avoid overfitting.

\noindent\textbf{Data augmentation}:  We utilize SpecAug, a cheap yet effective data augmentation method for spectrograms, has been introduced [31]. SpecAug randomly sets time-frequency regions to zero within an input log-Mel spectrogram with $D$ (here 64) number of frequency bins and $T$ frames. Time modification is applied by masking $y_t$ times $x_t$ consecutive time frames, where $x_t$ is chosen to be uniformly distributed between [ $t_0$, $t_0$ + $x_t$ ] and $t_0$ is uniformly distributed. Frequency modification is applied by masking $y_f$ times consecutive frequency bins[ $f_0$, $f_0$ + $x_f$ ), where randomly chosen from a uniform distribution in the range and $f_0$ is uniformly chosen. For all experiments we set $y_t$ = 2, $y_f$ = 2.

\subsection{Performance on ESC-50 and noiseESC-50}
Experimental results are shown in Table~\ref{tab:1}. Our model outperforms the previous model on two datasets. As shown in Table~\ref{tab:1}, the absolute improvement of our best model (TA+TPN+DA (temporal\&channel)) over published SOTA (TPN) is +6.8\% in \textit{5-way 1-shot}, +8.4\% in \textit{10-way 1-shot} on ESC-50. On noiseESC-50, +5.9\% in \textit{5-way 1-shot} and +7.8\% in \textit{10-way 1-shot}.
At the same time, on ESC-50, we also notice that the performance gains slightly improvement than the SOTA, +0.1\% in \textit{10-way 5-shot}, 0.3\% in \textit{5-way 5-shot} on ESC-50, +3.8\% in \textit{10-way 5-shot}, +1.0\% in \textit{5-way 5-shot} on noiseESC-50.
Obviously, our model gains more improvement in 1-shot setting. Two metric-based methods can be continuously improved by integrating TA.
Specifically, the experiment results of TPN on two datasets is produced by ourself. All results are averaged over 1000 Meta-train $\&$ Meta-test episodes.

From the statistics of Table~\ref{tab:1}, data augmentation 
and temporal\&channel attention mechanism are also contributive. 
The data augmentation strategy could continuously improve the effect of various few-shot models. The enhancement is more obvious of TA+TPN, 2.7\% for 5-way 1-shot and 2.0\% for 10-way 1-shot. The improvement brought by data augmentation and novel attention mechanism illustrate that the performance of baseline methods is severely underestimated.

\subsection{Analysis of experimental results}
\begin{figure}[t]
	\centering
	\includegraphics[width=\linewidth]{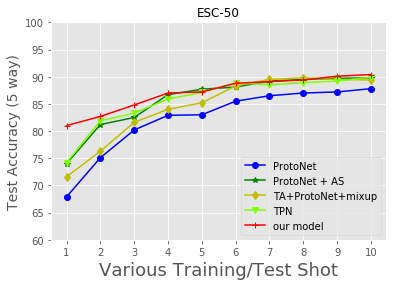}
	\caption{5-way performance on ESC-50 with various training/test shots}
	\label{fig:fig_exp_1}
\end{figure}

First, the temporal\&channel attention can significantly improve the sound's representation, thus promote model's performances. It is acknowledged that
temporal\&channel attention and data augmentation strategy could reduce the intra-class variation~\cite{dataaug}.
With regarding the noise, the performance on ESC-50 is inferior to noiseESC-50. 
In addition, another significant observation is that 5-shot is less significantly improved than 1-shot.
For example, in 5-way of ESC-50, the performance of our model over published state-of-the-art is 0.3\% for 5-shot but 6.8\% for 1-shot. 
Moreover, the margin of the various model decreases with the increasing of shots is because more labeled data are used. The superiority of task-adaptive module and other modules will be decreased when more labeled data are available. 
In this paper, We also make detail experiments (5-way k-shot $k \in \{1,2,...,10\}$) of various model, and the results are presented in Figure~\ref{fig:fig_exp_1} and Figure 5, which verify the viewpoints above.
This gives a qualitative comparison of the result of the prototypical network with and without attentional similarity for 3 query examples from the noiseESC-50 test set. Those picked by the model without attentional similarity (marked in cyan) do not share the same class as the queries; they are picked possibly because both the query and the picked one have a long silence. In contrast, the model with attentional similarity finds correct matches (marked in red).

As this sample shows, the TPN model is indeed capable of sound localization, specifically for events with a duration information, such as “gun shot”. However, it seems to struggle with longer events, such as “dog bark,” at which it exhibits a peaking behavior, chunking the event into small pieces.
On the contrary, TA can predict and localize both short and long events for this sample. Specifically, TA + TPN was unable to notice the short event, yet predicted its presence. We believe that this is due to the low time-resolution could skip over the presence of short events.

\begin{figure}[t]
	\centering
	\includegraphics[width=\linewidth]{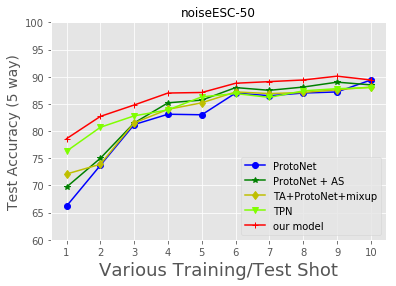}
	\caption{5-way performance on noiseESC-50 with various training/test shots}
	\label{fig:fig_exp_2}
\end{figure}

\begin{table}[t]
	\caption{The result of few-shot sound detection in domain mismatch. The AUC (Area Under Curve) is used for evaluation.}
	\label{tab:2}
	\centering
	\resizebox{1.0\linewidth}{24mm}{
		\begin{tabular}{c|c|c|c|c|c}
			\toprule[1.5pt]
			\multicolumn{2}{c|}{\multirow{2}{*}{Model}} & \multicolumn{2}{c|}{1-shot} &  \multicolumn{2}{c}{5-shot} \\ \cline{3-6} 
			\multicolumn{2}{c|}{}  & Music & Animal  & Music & Animal \\ \midrule
			\multicolumn{2}{c|}{ProtoNet\cite{protoNet}}  & ~~~0.712~~~  &  ~~~0.644~~~ & ~~~0.824~~~ & ~~~0.729~~~  \\ \midrule
			\multicolumn{2}{c|}{ProtoNet+AS\cite{attentionSimilarity}} & 0.736 & 0.677  & 0.839   & 0.750 \\ \midrule
			\multicolumn{2}{c|}{TPN\cite{TPN}}  & 0.747  & 0.685  & 0.843  & 0.748  \\
			\midrule \midrule
			\multicolumn{2}{c|}{\textbf{\textit{ProtoNet+DA}}}  & 0.752 & 0.680  & 0.836  & 0.739  \\ 
			\midrule
			\multicolumn{2}{c|}{\textbf{\textit{TA+ProtoNet}}}  & 0.759 & 0.691  & 0.851  & 0.763  \\ 
			\midrule
			\multicolumn{2}{c|}{\textbf{\textit{TA+ProtoNet+DA}}}  & 0.762 & 0.694  & 0.854  & 0.761  \\ 
			\midrule
			\multicolumn{2}{c|}{\textbf{\textit{TA+ProtoNet+DA(temporal\&channel)}}}  & \textbf{0.779} & \textbf{0.705}  & \textbf{0.857}  & \textbf{0.768}  \\ 
			\midrule
			\multicolumn{2}{c|}{\textbf{\textit{TPN+DA}}}  & 0.755 & 0.693  & 0.848  & 0.751  \\ 
			\midrule
			\multicolumn{2}{c|}{\textbf{\textit{TA+TPN}}}  & 0.762 & 0.691  & 0.854  & 0.753  \\ 
			\midrule
			\multicolumn{2}{c|}{\textbf{\textit{TA+TPN + DA}}}  & 0.768 & 0.696  & 0.850  & 0.754  \\ 
			\midrule
			\multicolumn{2}{c|}{\textbf{\textit{TA+TPN+DA(temporal\&channel)}}}  & \textbf{0.779} & \textbf{0.705}  & \textbf{0.857}  & \textbf{0.761}  \\ 
			\bottomrule[1.5pt]
		\end{tabular}
	}
\end{table}

\subsection{Analysis of domain mismatch}

During this subsection, we follow the setting created by~\cite{fewshotdomain}. While the current evaluation focus on recognizing novel class with limited training examples, these novel classes are sampled from the same domain. So, we follow the experiments from~\cite{fewshotdomain}, such a out-of-domain testing could display the ability of few-shot learning methods to generalize~\cite{domainad,domain_ad}. Following the previous setup, the selected AudioSet~\cite{audioSet} with 99 events for meta-train, meta-validation with 21 events and meta-test with 21 events. In addition, the pre-processing are same as the previous description about the setup of~\cite{attentionSimilarity}. The number of ways is set to 5, and we rerun the experiments: ProtoNet, ProtoNet+AS and TPN. For the domain mismatch setting, we experiments "Music" and "Animals" domain as same as~\cite{fewshotdomain}. The events and associated audios from the two domains are removed from train set.

Results on ProtoNet are slightly better than those on TPN, which is mainly because in Audioset music sounds are common and some patterns exist in other sounds as well such as bell ringing. On the other hand animal sounds such as “dog growling” rarely resemble other sounds. The gain of meta-learning approaches over supervised baselines are diminished due to domain mismatch. This implies the potential overfitting issue in meta-learning. Meta-learning models learn to utilize the correlation between classes. However, if all training classes come from one domain, the model tends to overfit to that domain and performance on new domain would drop.

\section{Conclusion}
In this paper, we proposed a task-adaptive module for few-shot sound event detection. The module contains a task-adaptive extractor (TAE) and a projector. By considering all support examples at same time, TAE could extract the inner-class commonality and the projector could find cross-class characteristic features. Besides, the data augmentation strategy and the novel attention mechanism could further improve model’s performance. We demonstrated that it significantly improved accuracy on two benchmarks (ESC-50 and noiseESC-50), achieving state-of-the-art performance. In addition, we compared with several baselines under the experimental domain mismatch setting, our model could also gain improvement over baselines.

In future work we plan to further analyze why adaptive training of current state-of-the-art models does not yield substantial improvements in the multi-shots setting. We would also like to devise a method that will perform better in fast adaptation and unsupervised adaptation. Finally, we plan to evaluate our method on more challenge tasks such as domain adaptation or language adaptation of ASR models.

\section*{Acknowledgment}
This paper is supported by the Key Research and Development Program of Guangdong Province under grant No.2021B0101400003. Corresponding author is Jianzong Wang from Ping An Technology (Shenzhen) Co., Ltd (jzwang@188.com).

\vfill
\pagebreak
\bibliographystyle{IEEEtran}
\bibliography{strings,refs}

\begin{thebibliography}{10}
\providecommand{\url}[1]{#1}
\csname url@samestyle\endcsname
\providecommand{\newblock}{\relax}
\providecommand{\bibinfo}[2]{#2}
\providecommand{\BIBentrySTDinterwordspacing}{\spaceskip=0pt\relax}
\providecommand{\BIBentryALTinterwordstretchfactor}{4}
\providecommand{\BIBentryALTinterwordspacing}{\spaceskip=\fontdimen2\font plus
\BIBentryALTinterwordstretchfactor\fontdimen3\font minus
  \fontdimen4\font\relax}
\providecommand{\BIBforeignlanguage}[2]{{%
\expandafter\ifx\csname l@#1\endcsname\relax
\typeout{** WARNING: IEEEtran.bst: No hyphenation pattern has been}%
\typeout{** loaded for the language `#1'. Using the pattern for}%
\typeout{** the default language instead.}%
\else
\language=\csname l@#1\endcsname
\fi
#2}}
\providecommand{\BIBdecl}{\relax}
\BIBdecl

\bibitem{TUTDatatbase}
A.~Mesaros, T.~Heittola, and T.~Virtanen, ``{TUT} database for acoustic scene
  classification and sound event detection,'' in \emph{Proceedings of
  {EUSIPCO}}, 2016, pp. 1128--1132.

\bibitem{r}
H.~Dinkel, M.~Wu, and K.~Yu, ``Towards duration robust weakly supervised sound
  event detection,'' \emph{IEEE/ACM Transactions on Audio, Speech, and Language
  Processing}, vol.~29, pp. 887--900, 2021.

\bibitem{fed}
Z.~Hong, J.~Wang, X.~Qu, J.~Liu, C.~Zhao, and J.~Xiao, ``{Federated Learning
  with Dynamic Transformer for Text to Speech},'' in \emph{Proc. Interspeech
  2021}, 2021, pp. 3590--3594.

\bibitem{g2p}
C.~Zhao, J.~Wang, X.~Qu, H.~Wang, and J.~Xiao, ``r-g2p: Evaluating and
  enhancing robustness of grapheme to phoneme conversion by controlled noise
  introducing and contextual information incorporation,'' in \emph{ICASSP
  2022-2022 IEEE International Conference on Acoustics, Speech and Signal
  Processing (ICASSP)}.\hskip 1em plus 0.5em minus 0.4em\relax IEEE, 2022, pp.
  6197--6201.

\bibitem{sed1}
Y.~Chen, H.~Dinkel, M.~Wu, and K.~Yu, ``Voice activity detection in the wild
  via weakly supervised sound event detection.'' in \emph{Proc. Interspeech
  2021}, 2020, pp. 3665--3669.

\bibitem{sed2}
S.~G. Upadhyay, B.~Su, and C.~Lee, ``Attentive convolutional recurrent neural
  network using phoneme-level acoustic representation for rare sound event
  detection,'' in \emph{Proceedings of {Interspeech}}, H.~Meng, B.~Xu, and
  T.~F. Zheng, Eds.\hskip 1em plus 0.5em minus 0.4em\relax {ISCA}, 2020, pp.
  3102--3106.

\bibitem{se}
W.~Xia and K.~Koishida, ``{Sound Event Detection in Multichannel Audio Using
  Convolutional Time-Frequency-Channel Squeeze and Excitation},'' in
  \emph{Proc. Interspeech 2019}, 2019, pp. 3629--3633.

\bibitem{5}
Y.~Tang, J.~Wang, X.~Qu, and J.~Xiao, ``Contrastive learning for improving
  end-to-end speaker verification,'' in \emph{2021 International Joint
  Conference on Neural Networks (IJCNN)}.\hskip 1em plus 0.5em minus
  0.4em\relax IEEE, 2021, pp. 1--7.

\bibitem{6}
Z.~Hong, J.~Wang, W.~Wei, J.~Liu, X.~Qu, B.~Chen, Z.~Wei, and J.~Xiao, ``When
  hearing the voice, who will come to your mind,'' in \emph{2021 International
  Joint Conference on Neural Networks (IJCNN)}.\hskip 1em plus 0.5em minus
  0.4em\relax IEEE, 2021, pp. 1--6.

\bibitem{3}
N.~Zhang, J.~Wang, W.~Wei, X.~Qu, N.~Cheng, and J.~Xiao, ``Cacnet: Cube
  attentional cnn for automatic speech recognition,'' in \emph{2021
  International Joint Conference on Neural Networks (IJCNN)}.\hskip 1em plus
  0.5em minus 0.4em\relax IEEE, 2021, pp. 1--7.

\bibitem{autoTagging}
J.~Liu and Y.~Yang, ``Event localization in music auto-tagging,'' in
  \emph{Proceedings of {ACMMM}}, 2016, pp. 1048--1057.

\bibitem{2}
X.~Qu, J.~Wang, and J.~Xiao, ``Evolutionary algorithm enhanced neural
  architecture search for text-independent speaker verification,'' \emph{Proc.
  Interspeech 2020}, pp. 961--965, 2020.

\bibitem{audiofew}
J.~Pons, J.~Serr{\`{a}}, and X.~Serra, ``Training neural audio classifiers with
  few data,'' in \emph{Proceedings of {ICASSP}}.\hskip 1em plus 0.5em minus
  0.4em\relax {IEEE}, 2019, pp. 16--20.

\bibitem{song}
H.~Song, J.~Han, S.~Deng, and Z.~Du, ``{Acoustic Scene Classification by
  Implicitly Identifying Distinct Sound Events},'' in \emph{Proc. Interspeech
  2019}, 2019, pp. 3860--3864.

\bibitem{park}
I.~Park and H.~K. Kim, ``{Two-Stage Polyphonic Sound Event Detection Based on
  Faster R-CNN-LSTM with Multi-Token Connectionist Temporal Classification},''
  in \emph{Proc. Interspeech 2020}, 2020, pp. 856--860.

\bibitem{few-shot-sound-detection}
Y.~Wang, J.~Salamon, N.~J. Bryan, and J.~P. Bello, ``Few-shot sound event
  detection,'' in \emph{Proceedings of {ICASSP}}.\hskip 1em plus 0.5em minus
  0.4em\relax {IEEE}, 2020, pp. 81--85.

\bibitem{attentionSimilarity}
S.~Chou, K.~Cheng, J.~R. Jang, and Y.~Yang, ``Learning to match transient sound
  events using attentional similarity for few-shot sound recognition,'' in
  \emph{Proceedings of {ICASSP}}, 2019, pp. 26--30.

\bibitem{rare}
W.~Wang, C.-C. Kao, and C.~Wang, ``A simple model for detection of rare sound
  events,'' in \emph{Proc. Interspeech 2018}, 2018, pp. 1344--1348.

\bibitem{few-att-gnn}
S.~Zhang, Y.~Qin, K.~Sun, and Y.~Lin, ``Few-shot audio classification with
  attentional graph neural networks,'' in \emph{Proceedings of {Interspeech}},
  G.~Kubin and Z.~Kacic, Eds.\hskip 1em plus 0.5em minus 0.4em\relax {ISCA},
  2019, pp. 3649--3653.

\bibitem{fewShotGNN}
V.~G. Satorras and J.~B. Estrach, ``Few-shot learning with graph neural
  networks,'' in \emph{Proceedings of {ICLR}}, 2018.

\bibitem{TPN}
Y.~Liu, J.~Lee, M.~Park, S.~Kim, E.~Yang, S.~J. Hwang, and Y.~Yang, ``Learning
  to propagate labels: Transductive propagation network for few-shot
  learning,'' in \emph{Proceedings of {ICLR}}, 2019.

\bibitem{MatchNet}
O.~Vinyals, C.~Blundell, T.~Lillicrap, K.~Kavukcuoglu, and D.~Wierstra,
  ``Matching networks for one shot learning,'' in \emph{Proceedings of {NIPS}},
  2016, pp. 3630--3638.

\bibitem{protoNet}
J.~Snell, K.~Swersky, and R.~S. Zemel, ``Prototypical networks for few-shot
  learning,'' in \emph{Proceedings of {NIPS}}, 2017, pp. 4077--4087.

\bibitem{relationNet}
F.~Sung, Y.~Yang, L.~Zhang, T.~Xiang, P.~H.~S. Torr, and T.~M. Hospedales,
  ``Learning to compare: Relation network for few-shot learning,'' in
  \emph{Proceedings of {CVPR}}, 2018, pp. 1199--1208.

\bibitem{CloserLook}
W.~Chen, Y.~Liu, Z.~Kira, Y.~F. Wang, and J.~Huang, ``A closer look at few-shot
  classification,'' in \emph{Proceedings of {ICLR}}.\hskip 1em plus 0.5em minus
  0.4em\relax OpenReview.net, 2019.

\bibitem{fewgnn}
V.~G. Satorras and J.~B. Estrach, ``Few-shot learning with graph neural
  networks,'' in \emph{Proceedings of {ICLR}}.\hskip 1em plus 0.5em minus
  0.4em\relax OpenReview.net, 2018.

\bibitem{maml}
C.~Finn, P.~Abbeel, and S.~Levine, ``Model-agnostic meta-learning for fast
  adaptation of deep networks,'' in \emph{Proceedings of {ICML}}, 2017, pp.
  1126--1135.

\bibitem{opt1}
S.~Ravi and H.~Larochelle, ``Optimization as a model for few-shot learning,''
  in \emph{Proceedings of {ICLR}}.\hskip 1em plus 0.5em minus 0.4em\relax
  OpenReview.net, 2017.

\bibitem{opt2}
A.~Nichol, J.~Achiam, and J.~Schulman, ``On first-order meta-learning
  algorithms,'' \emph{CoRR}, vol. abs/1803.02999, 2018.

\bibitem{koch2015siamese}
G.~Koch, R.~Zemel, and R.~Salakhutdinov, ``Siamese neural networks for one-shot
  image recognition,'' in \emph{ICML deep learning workshop}, vol.~2, 2015.

\bibitem{TSVM}
T.~Joachims, ``Transductive inference for text classification using support
  vector machines,'' in \emph{Proceedings of {ICML}}, 1999, pp. 200--209.

\bibitem{TPNzero}
Y.~Fu, T.~M. Hospedales, T.~Xiang, and S.~Gong, ``Transductive multi-view
  zero-shot learning,'' \emph{{IEEE} Trans. PAMI.}, vol.~37, no.~11, pp.
  2332--2345, 2015.

\bibitem{cvpr}
M.~Boudiaf, H.~Kervadec, Z.~I. Masud, P.~Piantanida, I.~Ben~Ayed, and J.~Dolz,
  ``Few-shot segmentation without meta-learning: A good transductive inference
  is all you need?'' in \emph{Proceedings of the IEEE/CVF Conference on
  Computer Vision and Pattern Recognition}, 2021, pp. 13\,979--13\,988.

\bibitem{timefrequnencymask}
D.~S. Park, W.~Chan, Y.~Zhang, C.~Chiu, B.~Zoph, E.~D. Cubuk, and Q.~V. Le,
  ``Specaugment: {A} simple data augmentation method for automatic speech
  recognition,'' in \emph{Proceedings of {Interspeech}}.\hskip 1em plus 0.5em
  minus 0.4em\relax {ISCA}, 2019, pp. 2613--2617.

\bibitem{dataa}
Y.~Chen and H.~Jin, ``{Rare Sound Event Detection Using Deep Learning and Data
  Augmentation},'' in \emph{Proc. Interspeech 2019}, 2019, pp. 619--623.

\bibitem{m}
M.~Boudiaf, I.~Ziko, J.~Rony, J.~Dolz, P.~Piantanida, and I.~Ben~Ayed,
  ``Information maximization for few-shot learning,'' \emph{Advances in Neural
  Information Processing Systems}, vol.~33, pp. 2445--2457, 2020.

\bibitem{dd}
V.~Morfi, D.~Stowell, V.~Lostanlen, A.~Strandburg-Peshkin, L.~Gill, H.~Pamula,
  D.~Benvent, I.~Nolasco, S.~Singh, S.~Sridhar \emph{et~al.}, ``Dcase 2021 task
  5: Few-shot bioacoustic event detection development set,'' 2021.

\bibitem{SimilairtyEmbeddingNetwork}
Y.~Huang, S.~Chou, and Y.~Yang, ``Generating music medleys via playing music
  puzzle games,'' in \emph{Proceedings of {AAAI}}, 2018, pp. 2281--2288.

\bibitem{librosa}
P.~Raguraman, M.~Ramasundaram, and M.~Vijayan, ``Librosa based assessment tool
  for music information retrieval systems,'' in \emph{Proceedings of {MIPR}},
  2019, pp. 109--114.

\bibitem{dataaug}
X.~Song, Z.~Wu, Y.~Huang, D.~Su, and H.~Meng, ``Specswap: {A} simple data
  augmentation method for end-to-end speech recognition,'' in \emph{Proceedings
  of {Interspeech}}, H.~Meng, B.~Xu, and T.~F. Zheng, Eds.\hskip 1em plus 0.5em
  minus 0.4em\relax {ISCA}, 2020, pp. 581--585.

\bibitem{fewshotdomain}
B.~Shi, M.~Sun, K.~C. Puvvada, C.~Kao, S.~Matsoukas, and C.~Wang, ``Few-shot
  acoustic event detection via meta-learning,'' in \emph{Proceedings of
  {ICASSP}}.\hskip 1em plus 0.5em minus 0.4em\relax IEEE, 2020, pp. 76--80.

\bibitem{domainad}
Y.~Ganin, E.~Ustinova, H.~Ajakan, P.~Germain, H.~Larochelle, F.~Laviolette,
  M.~Marchand, and V.~S. Lempitsky, ``Domain-adversarial training of neural
  networks,'' in \emph{Domain Adaptation in Computer Vision Applications}, ser.
  Advances in CVPR, G.~Csurka, Ed.\hskip 1em plus 0.5em minus 0.4em\relax
  Springer, 2017, pp. 189--209.

\bibitem{domain_ad}
S.~Motiian, Q.~Jones, S.~M. Iranmanesh, and G.~Doretto, ``Few-shot adversarial
  domain adaptation,'' in \emph{Proceedings of {NeurIPS}}, I.~Guyon, U.~von
  Luxburg, S.~Bengio, H.~M. Wallach, R.~Fergus, S.~V.~N. Vishwanathan, and
  R.~Garnett, Eds., 2017, pp. 6670--6680.

\bibitem{audioSet}
J.~F. Gemmeke, D.~P.~W. Ellis, D.~Freedman, A.~Jansen, W.~Lawrence, R.~C.
  Moore, M.~Plakal, and M.~Ritter, ``Audio set: An ontology and human-labeled
  dataset for audio events,'' in \emph{Proceedings of {ICASSP}}, 2017, pp.
  776--780.

\end{thebibliography}

\end{document}